\documentclass[12pt]{article}

\usepackage{amsmath,amssymb}
\usepackage{graphicx}

\makeatletter

\renewcommand\section{\@startsection{section}{1}{\z@}
                                   {-3.5ex \@plus -1ex \@minus -.2ex}
                                   {2.3ex \@plus .2ex}
                                   {\normalfont\large\bfseries}}
\renewcommand\subsection{\@startsection{subsection}{2}{\z@}
                                   {-3.25ex\@plus -1ex \@minus -.2ex}
                                   {1.5ex \@plus .2ex}
                                   {\normalfont\normalsize\bfseries}}
\renewcommand\subsubsection{\@startsection{subsubsection}{3}{\z@}
                                   {-3.25ex\@plus -1ex \@minus -.2ex}
                                   {1.5ex \@plus .2ex}
                                   {\normalfont\normalsize\bfseries}}
\renewcommand\paragraph{\@startsection{paragraph}{4}{\z@}
                                   {3.25ex \@plus1ex \@minus.2ex}
                                   {-1em}
                                   {\normalfont\normalsize\bfseries}}

\makeatother

\newcommand{\be}{\begin{equation}}
\newcommand{\ee}{\end{equation}}
\newcommand{\bea}{\begin{eqnarray}}
\newcommand{\eea}{\end{eqnarray}}
\newcommand{\ba}{\begin{array}}
\newcommand{\ea}{\end{array}}

\newcommand{\id}{\hbox{1\kern-.27em l}}
\newcommand{\ZZ}{\mathbb{Z}}

\newcommand{\half}{ {\textstyle \frac{1}{2}  } }

\newcommand{\ga}{\gamma}
\newcommand{\Ga}{\Gamma}

\newcommand{\ep}{\epsilon}
\newcommand{\si}{\sigma}

\newcommand{\cN}{\mathcal{N}}

\newcommand{\rar}{\rightarrow}
\newcommand{\non}{\nonumber}

\newcommand{\SU}{\mathrm{SU}}
\newcommand{\SO}{\mathrm{SO}}
\newcommand{\SL}{\mathrm{SL}}
\newcommand{\Sp}{\mathrm{Sp}}
\newcommand{\su}{\mathrm{su}}
\newcommand{\so}{\mathrm{so}}
\newcommand{\spl}{\mathrm{sp}}
\newcommand{\Spin}{\mathrm{Spin}}
\newcommand{\Pin}{\mathrm{Pin}}

\newcommand{\ts}{\textstyle}

\newcommand{\beq}{\begin{equation}}
\newcommand{\eeq}{\end{equation}}
\newcommand{\column}[3]{\left( \begin{array}{c} #1 \cr #2 \cr #3 \end{array} \right)}
\newcommand{\col}[3]{\left[ \begin{array}{c} #1 \cr #2 \cr #3 \end{array} \right]}

\begin{document}
\begin{center}

\vspace*{5mm}
{\Large\sf  Bound states in $\cN = 4$ SYM on $T^3$\,:\\[0.5mm]
$\Spin (2 n)$ and the exceptional groups}

\vspace*{5mm}
{\large M{\aa}ns Henningson, Niclas Wyllard}

\vspace*{5mm}
Department of Fundamental Physics\\
Chalmers University of Technology\\
S-412 96 G\"oteborg, Sweden\\[3mm]
{\tt mans,wyllard@fy.chalmers.se}     
     
\vspace*{5mm}{\bf Abstract} 

\end{center}

\noindent The low energy spectrum of (3+1)-dimensional $\cN=4$ supersymmetric Yang-Mills theory on a spatial three-torus contains a certain number of bound states, characterized by their discrete abelian magnetic and electric 't~Hooft fluxes. At weak coupling, the wave-functions of these states are supported near points in the moduli space of flat connections where the unbroken gauge group is semi-simple. The number of such states is related to the number of normalizable bound states at threshold in the supersymmetric matrix quantum mechanics with 16 supercharges based on this unbroken group. Mathematically, the determination of the spectrum relies on the classification of almost  commuting triples with semi-simple centralizers. We complete the work begun in a previous paper, by computing the spectrum of bound states in theories based on the even-dimensional spin groups and the exceptional groups. The results satisfy the constraints of $S$-duality in a rather non-trivial way.

\setcounter{equation}{0}
\section{Introduction}

Recently, we initiated a study of the low-energy spectrum of (3+1)-dimensional ${\cN=4}$ supersymmetric Yang-Mills theory on $\mathbb{R}{\times}T^3$ \cite{Henningson:2007}, by considering the cases based on $\SU(n)$, $\Spin(2n{+}1)$ and $\Sp(2n)$. In this paper we consider the remaining cases, i.e.~the even-dimensional spin groups and the exceptional groups. 

Since the $\cN=4$ Yang-Mills theory contains only adjoint fields and is formulated on a spatial three torus, its states can be characterized by the discrete abelian magnetic and electric 't Hooft fluxes $m$ and $e$ \cite{'tHooft:1979}. We have
\bea
m & \in & M \simeq H^2 (T^3, C) \cr
e & \in & E \simeq \mathrm{Hom} (H^1 (T^3, C), U (1)) \, ,
\eea
where $C$ is the center of the simply connected cover $G$ of the gauge group. By a choice of three one-cycles generating $H_1 (T^3, \mathbb Z) \simeq (\mathbb Z)^3$, we may identify $m$ and $e$ with triples valued in $C$:
\bea
m & = & (m_{23},m_{31},m_{12}) \in C^3 \cr
e & = & (e_1,e_2,e_3) \in C^3 .
\eea  
Not all combinations of $m$ and $e$ may appear in a gauge theory, though: If the gauge group is a simply connected group $G$, we have $m = 0$ and $e$ can be arbitrary, whereas if the gauge group is $G / C$, we have $m$ arbitrary and $e = 0$. There are also intermediate cases where the gauge group is given by the quotient of $G$ by a non-trivial proper subgroup of its center $C$. We will, however, be slightly more general and consider all combinations of $m \in C^3$ and $e \in C^3$.

The wave-functions of low-energy states are supported near flat connections on the gauge bundle. Such a connection is classified by its holonomies $U_i$, $i = 1, 2, 3$ along the non-trivial cycles of $T^3$. The holonomies commute when regarded as elements of the gauge group, but when lifted to the simply connected cover $G$ they need only be almost commuting in the sense that
\be
U_i U_j = m_{ij} U_j U_i \, ,
\ee
where the $m_{ij}$ are the components of the magnetic 't Hooft flux. Gauge transformations that are continuously connected to the identity act by simultaneous conjugation on the $U_i$ by an element of the gauge group.  At a generic point in the moduli space of gauge equivalence classes of such almost commuting triples, the gauge group is broken to a subgroup that contains an abelian factor. The corresponding quantum states are then not normalizable because of the abelian scalar fields. But points in the moduli space where the unbroken gauge group is semi-simple gives rise to bound states of exactly zero energy; the theory in a neighborhood of such a point may be modelled by supersymmetric matrix quantum mechanics with 16 supercharges based on the Lie algebra of the unbroken gauge group, and such quantum mechanical theories are believed to have bound states~\cite{Witten:1995,Kac:1999b}, which in turn lead to bound states in the $\cN=4$ Yang-Mills theory on $\mathbb{R}{\times}T^3$. (This argument is best carried out in the weak coupling limit, but it is expected that the spectrum of low-energy states is invariant under continuous deformations of the theory, allowing for an interpolation between the strong and weak coupling regimes.) Diagonalizing the action of large gauge transformations, which act by multiplication of the holonomies by elements of the center of the gauge group, gives a spectrum of values of the electric 't Hooft flux $e$.

The spectrum of low energy states of the Yang-Mills theory should be invariant under $S$-duality \cite{Montonen:1977}, which (using multiplicative notation) acts on the electric and magnetic 't Hooft fluxes as follows:
\be \label{S}  \ba{llll}
T: & (m, e) \mapsto (m, em) & {\mathrm as}  & \tau \mapsto \tau + 1 \\
S: &  (m, e) \mapsto (e, m^{-1}) & {\mathrm as} & \tau \mapsto - 1 / \tau \, .
\ea \ee
In our previous paper, we showed that this gives a overdetermined set of equations for the dimensions of the spaces of bound states in quantum-mechanical models based on the classical matrix Lie algebras. (In that paper, in addition to normalizable states, we also considered continua of states of arbitrarily low energy, but this yields essentially no further information.) In the present paper, we will get further checks on these results and also (almost) determine the number of bound states for the quantum mechanical models based on the exceptional Lie algebras. In fact, $S$-duality implies that the spaces $V_s$ of normalizable states in supersymmetric quantum mechanics based on the Lie algebra $s$ must obey
\bea
\dim V_s = \left\{
\begin{array}{ll}
\mbox{$1$} & \mbox{for } s \simeq \su (n) \cr
\mbox{\# of partitions of $n$ into} \cr
\mbox{distinct odd parts} & \mbox{for } s \simeq \so (n) \cr
\mbox{\# of partitions of $n$ into} \cr
\mbox{distinct parts} & \mbox{for } s \simeq \spl (2n) \cr
\mbox{$2 + \Delta_{G_2}$} & \mbox{for } s \simeq G_2 \cr
\mbox{$4 + \Delta_{F_4}$} & \mbox{for } s \simeq F_4 \cr
\mbox{$3 + \Delta_{G_2}$} & \mbox{for } s \simeq E_6 \cr
\mbox{$6 + \Delta_{F_4}$} & \mbox{for } s \simeq E_7 \cr
\mbox{$11 + \Delta_{E_8}$} & \mbox{for } s \simeq E_8
\end{array}
\right.
\eea
for some undetermined integers $\Delta_{G_2}$, $\Delta_{F_4}$, and $\Delta_{E_8}$. We would like to remark that if one takes $\Delta_{G_2} = \Delta_{F_4} = \Delta_{E_8} = 0$, our results agree with those obtained by Kac and Smilga \cite{Kac:1999b} for the number of ground states in the mass-deformed ($N = 1^*$) theories. In the following we will for simplicity assume that these are indeed the correct values; it is trivial to insert other choices of $\Delta_{G_2}$, $\Delta_{F_4}$, and $\Delta_{E_8}$ in our formulas if one wishes to do so.
 
In the next section, we consider the cases based on the even-dimensional spin groups $\Spin (2 n)$ (where one needs to distinguish the cases where the dimension $2 n$ equals $0$ or $2$ modulo $4$), and in the last section we  consider the exceptional groups $G_2$, $F_4$, $E_6$, $E_7$, and $E_8$. The case by case analysis is rather tedious and we will mostly content ourselves with giving the results. More details can be found in our previous paper \cite{Henningson:2007}. Very useful background material from mathematics and physics can also be found respectively in \cite{Borel:1999} and \cite{Witten:2000}. Finally, we would like to remark that we expect a more intuitive explanation of the findings of this and our previous paper to be forthcoming.

\setcounter{equation}{0}
\section{The even-dimensional spin groups}

The centre of $G=\Spin(2n)$ is $C=\{\id,-\id,\Ga,-\Ga\}$, where $\Ga = \ga_1\cdots \ga_{2n}$. Since $\Ga^2 = (-\id)^n$, the centre is 
isomorphic to $\ZZ_4$ when $n$ is odd, i.e.~for $\Spin(4k+2)$, and is isomorphic to $\ZZ_2{\times}\ZZ_2$ when $n$ is even, i.e.~for $\Spin(4k)$. (For all $n$, $\{\id,-\id\}$ is the $\ZZ_2$ subgroup of the centre which upon quotioning $G$ by it gives the $\SO(2n)$ theory.)
Using the same notation for the elements of the centre of $\Spin(4k+2)$ and $\Spin(4k)$ allow us to treat some aspects of these two classes on a common footing. 
A slight drawback is that this notation does not stress the differences between $\Spin(4k{+}2)$ and $\Spin(4k)$.

As in \cite{Henningson:2007}, we wish to compute the generating function
\beq
f (m, e) = \sum_{n = 0}^\infty q^{2 n } \mathrm{mult}_{\Spin (2 n )}^0 (m, e) \,,
\eeq
where $\mathrm{mult}_{\Spin (2 n )}^0 (m, e)$ denotes the number of bound states in the $\Spin(2n)$ theory  with discrete 't Hooft fluxes $(m,e) \in C^3 \times C^3$.
Even though the formul\ae{} we obtain can be written in such a way that they are valid in both $\Spin(4k)$ and $\Spin(4k{+}2)$, we will treat these cases separately 
when it leads to increased clarity.

For $\Spin(4k)$, the equivalence classes of $m$ modulo the action of the  $\SL (3, \mathbb Z)$ mapping class group of $T^3$ may be represented by the following elements of $C^3$:
\beq
\begin{array}{crl}
\underline{m} & \underline{\#} & \qquad \underline{\mathrm{components}} \cr
(\id, \id, \id) &\qquad 1 & \qquad {\cal M}_{2 k} , \; {\cal M}_{2 k - 4} \cr
(\id, \id, -\id) & \qquad 7 & \qquad {\cal M}_{2 k - 2} , \; {\cal M}_{2 k - 2}^\prime \cr
(\id, \Ga, \Ga) &\qquad 7 & \qquad {\cal M}_{k} , \; {\cal M}_{k}^\prime, \;  {\cal M}_{k - 3} , \; {\cal M}_{k - 3}^\prime \cr
(\id, -\Ga, -\Ga) & \qquad 7 & \qquad {\cal M}_{k} , \; {\cal M}_{k}^\prime, \;  {\cal M}_{k - 3} , \; {\cal M}_{k - 3}^\prime \cr
(-\id, \Ga, \Ga) & \qquad 42 & \qquad {\cal M}_{k - 1} , \; {\cal M}_{k - 1}^\prime,  \; {\cal M}_{k - 2} , \; {\cal M}_{k - 2}^\prime ,
\end{array}
\eeq
where we have indicated the cardinality of the $\SL(3, \mathbb Z)$ orbit and the different components of the moduli space of flat connections with the rank of the unbroken subgroup as a subscript. The cases $m = (\id, \Ga, \Ga)$ and $m = (\id, -\Ga, -\Ga)$ are related by the automorphism that exchanges the two spinor representations.

For $\Spin(4k{+}2)$ we find instead:
\beq
\begin{array}{crl}
\underline{m} & \underline{\#} & \qquad \underline{\mathrm{components}} \cr
(\id, \id, \id) & \qquad 1 &\qquad {\cal M}_{2 k+1} , \; {\cal M}_{2 k - 3} \cr
(\id, \id, -\id) & \qquad 7 &\qquad {\cal M}_{2 k - 1} , \; {\cal M}_{2 k - 1}^\prime \cr
(\id, \Ga, \Ga) & \qquad 56 & \qquad {\cal M}^{(1)}_{k-1} , \; {\cal M}^{(2)}_{k-1}, \;  {\cal M}^{(3)}_{k - 1} , \; {\cal M}^{(4)}_{k - 1} .
\end{array}
\eeq

To describe the results, we define \cite{Henningson:2007,Kac:1999b} the generating functions for the number of bound states in $\so (n)$ and $\spl (2 n)$ quantum mechanics: 
\bea
P (q) & = & \sum_{n = 1}^\infty q^n \dim V_{\so (n)} = \prod_{k = 1}^\infty (1 + q^{2 k -1}) \cr
Q (q) & = & \sum_{n = 1}^\infty q^{2 n} \dim V_{\spl (2 n)} = \prod_{k = 1}^\infty (1 + q^{2 k}) .
\eea
We will also need the decomposition of $P (q)$ into its even and odd powers:
\bea
P_{\mathrm{even}} (q) & = & \frac{1}{2} \left(P (q) + P (-q) \right) \cr
P_{\mathrm{odd}} (q) & = & \frac{1}{2} \left(P (q) - P (-q) \right) .
\eea

\subsection{The components with vector structure}
From the above tables we see that the moduli spaces corresponding to the $\SL(3,\ZZ)$ orbits represented by $m=(\id,\id,\id)$ and $(\id,\id,-\id)$ can be given a common formulation for $\Spin(2n)$. 
For these values of $m$, it is always possible to embed the holonomies into a $\left[\Spin (l) {\times} \Spin (2n - l) \right] / {\sim}$ subgroup of $\Spin (2n)$, where the equivalence relation $\sim$ identifies the $-\id$ elements of the two factors.

As in \cite{Henningson:2007}, the part of the holonomies contained in the $\Spin(l)$ factor can be constructed from the following eight building blocks (which one may visualise as the corners of a cube):
\be \label{Spinfrac}
\column{\id}{\id}{\id}, \; \column{\ga}{\id}{\id}, \; \column{\id}{\ga}{\id}, \; \column{\id}{\id}{\ga}, \; \column{\ga}{\ga}{\id}, \; \column{\ga}{\id}{\ga}, \; \column{\id}{\ga}{\ga},  \; \column{\ga}{\ga}{\ga}.
\ee
Here $\ga$ denotes one of the usual gamma matrices $\ga_1, \ldots, \ga_l$. Not all combinations of the above 
building blocks give rise to holonomies that lie in $\Spin(l)$ (since each entry in (\ref{Spinfrac}) can be viewed as 
a $\Pin(1)$ element). In addition to the $l=1,3,5,7$ possibilities used 
in \cite{Henningson:2007} to describe the moduli spaces for $G=\Spin(2n{+}1)$ one 
can also have $l=0,4,8$.

\subsubsection{The $m = (\id, \id, \id)$ components}
On the ${\cal M}_n$ component, the holonomies are
\beq
\column{U_1}{U_2}{U_3} = \column{t_1}{t_2}{t_3} ,
\eeq
where the $t_i$ belong to a maximal torus $T^{n}$ of $\Spin (2n)$. 
An explicit representation is 
\be \label{torus}
t_i = \exp(\half \sum_{l\;\mathrm{odd}} \theta^l_i \ga_{l}\ga_{l+1} )
\ee
In this paper we are only interested in the points of the moduli space where the unbroken gauge group is semi-simple. Such gauge enhancement occurs at the points where 
all $\theta_i^l \in \{0,\pi\}$. At these points, the $t_i$ reduce to a product of 
$\id$ and $\ga_l\ga_{l+1}$ factors. The various possibilities can be seen as selecting 
an even number for each of the eight possibilities in (\ref{Spinfrac}). 
 
Enhanced $\bigoplus_{i=1}^{8} \so (2 n_i)$ symmetry occurs when
\beq \label{enhance}
\column{t_1}{t_2}{t_3} = \column{\ep_1}{\ep_2}{\ep_3} \column{\id}{\id}{\id}^{2 n_1} \ldots \column{\ga}{\ga}{\id}^{2 n_4} \column{\id}{\id}{\ga}^{2 n_5} \ldots \column{\ga}{\ga}{\ga}^{2 n_8} .
\eeq
Here $\ep_1$, $\ep_2$ and $\ep_3$ are sign factors. Depending on how many of the $n_i$'s are non-zero (i.e.~how many of the eight corner points of the cube are occupied), some or all of these may be removed by gauge transformations (see \cite{Henningson:2007}).

On the ${\cal M}_{n-4}$ component, the holonomies are
\beq
\column{U_1}{U_2}{U_3} = \left( \ba{cccccccc} \ga_1 &\ga_2 &\ga_3 &\ga_4 &\id &\id &\id & \id \\ \ga_1& \ga_2 &\id &\id& \ga_5 &\ga_6& \id &\id \\\ga_1 &\id &\ga_3 &\id& \ga_5& \id& \ga_7 &\id \ea \right) \column{t_1}{t_2}{t_3} ,
\eeq
where the $t_i$ belong to a maximal torus $T^{n-4}$ of $\Spin (2n - 8)$. 

When the $t_i$ take the form (\ref{enhance}), enhanced $\bigoplus_{i=1}^8 \so (2 n_i + 1)$ symmetry occurs. Since all eight corner points are occupied, all sign factors $\ep_i$ can be removed. This implies that the center element $-\id$ acts trivially, so there are no contributions for $e = (\id, \Ga, \Ga)$, $e = (\id, -\Ga, - \Ga)$, or $e = (- \id, \Ga, \Ga)$. (For $\Spin(4k+2)$ these values of $e$ are all related by $\SL(3,\ZZ)$; for $\Spin(4k)$ they represent distinct orbits.) But  $\Ga$ may have a non-trivial action (and acts in the same way as $-\Ga$). It acts trivially in all directions if all eight points on the cube are equally occupied, two combinations act trivially if the points within each of two parallel planes are equally occupied, and one combination acts trivially if the points within each of four parallel lines are equally occupied. Furthermore, the total number of states is easily seen to be $P_{\mathrm{odd}}^8 (q)$. In this way, one finds that the contribution for $e = (\id, \id, \id)$ is $\frac{1}{8} P_{\mathrm{odd}}^8 (q) + \frac{7}{8} P_{\mathrm{odd}}^4 (q^2)$, and the contribution for $e = (\id, \id, -\id)$ is $\frac{1}{8} P_{\mathrm{odd}}^8 (q) - \frac{1}{8} P_{\mathrm{odd}}^4 (q^2)$. Note that the terms with argument equal to $q^2$ only contribute in the $\Spin(4k)$ theories.

On the ${\cal M}_{n}$ component, depending on the number of occupied points, a number of relations between the signs $\ep_i = \pm$ may be imposed by conjugation (see \cite{Henningson:2007} for a discussion). Assume first that three independent sign relations may be imposed. The calculation is then analogous to the ${\cal M}_{n-4} ((\id, \id, \id))$ case, and yields the contribution $\frac{1}{8} P_{\mathrm{even}}^8 (q) + \frac{7}{8} P_{\mathrm{even}}^4 (q^2)$ for $e = (\id, \id, \id)$, and the contribution $\frac{1}{8} P_{\mathrm{even}}^8 (q) - \frac{1}{8} P_{\mathrm{even}}^4 (q^2)$ for $e = (\id, \id, -\id)$. In total, we thus get
\bea
\!\!\!\!\!\!f ((\id, \id, \id), (\id, \id, \id)) & \!\!= &\!\! \frac{1}{8} P_{\mathrm{odd}}^8 (q) + \frac{7}{8} P_{\mathrm{odd}}^4 (q^2) + \frac{1}{8} P_{\mathrm{even}}^8 (q) + \frac{7}{8} P_{\mathrm{even}}^4 (q^2) \cr
\!\!\!\!\!\!f ((\id, \id, \id), (\id, \id, -\id)) & \!\!= &\!\! \frac{1}{8} P_{\mathrm{odd}}^8 (q) - \frac{1}{8} P_{\mathrm{odd}}^4 (q^2) + \frac{1}{8} P_{\mathrm{even}}^8 (q) - \frac{1}{8} P_{\mathrm{even}}^4 (q^2) \,.
\eea
Including also the $\ep_i$ signs gives extra states, but these have $e$ taking one of the $56$ values where at least one component equals $\Ga$ or $-\Ga$. So far the above results are thus actually correct. One extra state occurs when only one of $2 \cdot 7$ possible planes is occupied. (Three extra states occur when only one line is occupied, but this case has already appeared within three of the single plane cases. Finally, seven extra states occur when only one point is occupied, but this has already occured within seven of the single plane cases.) Consider the two cases when the occupied plane is orthogonal to a given direction. There are a total of $P_{\mathrm{even}}^4 (q)$ extra  states for each of the values $\Ga$ and $-\Ga$ of the corresponding component of $e$. The element $\Ga$ may act non-trivially along the plane (and $-\Ga$ acts in the same way, unless we are in the case where only a line is occupied). Two such transformations act trivially if the whole plane is equally  occupied, and one acts trivially if the points within each of two parallel lines are equally occupied. In this way, one finds that
\bea
f ((\id, \id, \id), (\id, \Ga, \Ga)) & =  & \frac{1}{4} P_{\mathrm{even}}^4 (q) + \frac{3}{4} P_{\mathrm{even}}^2 (q^2) \cr
f ((\id, \id, \id), (\id, -\Ga, -\Ga)) & =  & \frac{1}{4} P_{\mathrm{even}}^4 (q) + \frac{3}{4} P_{\mathrm{even}}^2 (q^2) .
\eea
Using that the total number of states is $P_{\mathrm{even}}^8 (q) + 14 P_{\mathrm{even}}^4 (q)$, one finally finds that
\beq
f ((\id, \id, \id), (-\id, \Ga, \Ga)) = \frac{1}{4}  P_{\mathrm{even}}^4 (q) - \frac{1}{4} P_{\mathrm{even}}^2 (q^2) .
\eeq
Again, the terms with the argument $q^2$ only contribute in the $\Spin(4k)$ theory. This is 
in agreement with the fact that, in the $\Spin(4k{+}2)$ theory, the above three entries belong to the same $\SL(3,\ZZ)$ orbit.

\subsubsection{The $m = (\id, \id, -\id)$ components}
On the ${\cal M}_{n - 2}$ and ${\cal M}_{n - 2}^\prime$ components, the holonomies are
\beq
\column{U_1}{U_2}{U_3} =  \left( \ba{cccc} \ga_1 &\ga_2 &\id  &\id \\ \ga_1 &\id &\ga_3& \id \\\id &\id &\id &\id \ea \right) \column{t_1}{t_2}{t_3} ,
\eeq
and
\beq
\column{U_1}{U_2}{U_3} =  \left( \ba{cccc} \ga_1 &\ga_2 &\id  &\id\\ \ga_1 &\id &\ga_3 &\id\\\ga_1& \ga_2& \ga_3 &\ga_4 \ea \right) \column{t_1}{t_2}{t_3}
\eeq 
respectively, where the $t_i$ belong to a maximal torus $T^{n - 2}$ of $\Spin (2n - 4)$. The signs $\ep_1$ and $\ep_2$ can be removed by conjugation. Whether $\ep_3$ can be removed depends on the number of occupied points (see \cite{Henningson:2007}). 
Enhanced $\so (2 n_1 + 1) \oplus \ldots \oplus \so (2 n_4 + 1) \oplus \so (2 n_5) \oplus \ldots \oplus \so (2 n_8)$ and $\so (2 n_1) \oplus \ldots \oplus \so (2 n_4) \oplus \so (2 n_5+1) \oplus \ldots \oplus \so (2 n_8+1)$ symmetry respectively occurs when the $t_i$'s take the form (\ref{enhance}).

Assume first that the sign $\ep_3$ may be fixed. The total number of states on the two components is then $2 P_{\mathrm{even}}^4 (q) P_{\mathrm{odd}}^4 (q)$, and the center element $-\id$ acts trivially in all three directions. $\Ga$ (or equivalently $-\Ga$) always acts non-trivially in the 3-direction. It acts trivially in the 1- and 2-directions if the points within both the odd and even planes are equally occupied, and it acts trivially in one direction if the points along all lines in that direction are equally occupied. In this way, one finds that
\bea
\!\!\!\!\!\! f ((\id, \id, -\id), (\id, \id, \id)) & = & \frac{1}{4} P_{\mathrm{even}}^4 (q) P_{\mathrm{odd}}^4 (q) + \frac{3}{4} P_{\mathrm{even}}^2 (q^2) P_{\mathrm{odd}}^2 (q^2) \cr
\!\!\!\!\!\! f ((\id, \id, -\id), (\id, \id, -\id)) & = & \frac{1}{4} P_{\mathrm{even}}^4 (q) P_{\mathrm{odd}}^4 (q) + \frac{3}{4} P_{\mathrm{even}}^2 (q^2) P_{\mathrm{odd}}^2 (q^2) \cr
\!\!\!\!\!\! f ((\id, \id, -\id), (\id, -\id, \id)) & = & \frac{1}{4} P_{\mathrm{even}}^4 (q) P_{\mathrm{odd}}^4 (q) - \frac{1}{4} P_{\mathrm{even}}^2 (q^2) P_{\mathrm{odd}}^2 (q^2) \cr
\!\!\!\!\!\! f ((\id, \id, -\id), (\id, -\id, -\id)) & = & \frac{1}{4} P_{\mathrm{even}}^4 (q) P_{\mathrm{odd}}^4 (q) - \frac{1}{4} P_{\mathrm{even}}^2 (q^2) P_{\mathrm{odd}}^2 (q^2) .
\eea
Note that again the terms with $q^2$ argument only contribute in the $\Spin(4k)$ theory.  
When only the odd plane is occupied, the sign $\ep_3$ is relevant and gives $P_{\mathrm{odd}}^4 (q)$ extra states for each of the values $e_3 = \Ga$ and $e_3 = -\Ga$. The action of $\Ga$ (or equivalently $-\Ga$) in the 1- and 2-directions is as before. In this way, one finds that
\bea
 f ((\id, \id, -\id), (\id, \id, \Ga)) & = & \frac{1}{4} P_{\mathrm{odd}}^4 (q) + \frac{3}{4} P_{\mathrm{odd}}^2 (q^2) \cr
 f ((\id, \id, -\id), (\id, \id, -\Ga)) & = & \frac{1}{4} P_{\mathrm{odd}}^4 (q) +  \frac{3}{4} P_{\mathrm{odd}}^2 (q^2) \cr
 f ((\id, \id, -\id), (\id, -\id, \Ga)) & = & \frac{1}{4} P_{\mathrm{odd}}^4 (q) - \frac{1}{4} P_{\mathrm{odd}}^2 (q^2) \cr
 f ((\id, \id, -\id), (\id, -\id, -\Ga)) & = & \frac{1}{4} P_{\mathrm{odd}}^4 (q) - \frac{1}{4} P_{\mathrm{odd}}^2 (q^2) . 
\eea
Again the $q^2$ corrections only appear in the $\Spin(4k)$ theory as required by $\SL(3,\ZZ)$.

\subsection{The components without vector structure}
Next we turn to the remaining cases, i.e.~the choices of $m$ that involve at least one of $\Ga$ and/or $-\Ga$.  
For these values of $m$, it is always possible to embed the holonomies into 
a $\left[\Spin (2l) \times (\SU (2)_L \times \SU (2)_R)^{(n - l)/2} \right] / {\sim}$ subgroup of $\Spin (2n)$, where the equivalence relation identifies the element $-\id$ of $\Spin (2l)$ with the element $(-\id, -\id)$ of each $\SU (2)_L \times \SU (2)_R$ factor. When $n$ (and $l$) is even, the center of $\Spin (2n)$ is generated by $\Ga = (\Ga_\natural, (\id, -\id)^{(n-l)/2 })$, where $\Ga_\natural = \ga_1\cdots\ga_{2l}$ is a generator of the center of $\Spin (2l)$. For $n$ (and $l$) odd, one also needs $-\Ga = (-\Ga_\natural, (\id, -\id)^{(n-l)/2 }) $) to generate the full centre.

The part of the holonomies contained in the $\Spin(2l)$ factor can be constructed from six different building blocks, which can be taken to be e.g.:
\be \ba{ccc} \label{Spinfrac2}
\column{\ts \frac{1}{\sqrt{2}}(1+\ga_i\ga_j)}{\ga_j}{\ga_j}, &\qquad
\column{\ts \frac{1}{\sqrt{2}}(1+\ga_i\ga_j)}{\ga_j}{\ga_i}, &\qquad
\column{\ga_j}{\ts \frac{1}{\sqrt{2}}(\ga_i - \ga_j)}{\ts \frac{1}{\sqrt{2}}(\ga_i - \ga_j)},  \\
\column{\ga_j}{\ts \frac{1}{\sqrt{2}}(\ga_i - \ga_j)}{\ts \frac{1}{\sqrt{2}}(\ga_i + \ga_j)}, &\qquad
\column{\ga_j}{\ts \frac{1}{\sqrt{2}}(1-\ga_i\ga_j)}{\ts \frac{1}{\sqrt{2}}(1-\ga_i\ga_j)}, &\qquad
\column{\ga_j}{\ts \frac{1}{\sqrt{2}}(1-\ga_i\ga_j)}{\ts \frac{1}{\sqrt{2}}(1+\ga_i\ga_j)} . 
\ea \ee
Each entry can be viewed as an element in $\Pin(2)$ and not every combination corresponds to an element in $\Spin(2l)$ (i.e.~contains only even numbers of gamma matrices). It turns out that only $l=0,2,3,4,6$ are possible. Depending on which combinations are selected, one of the values $m=(\pm\id,\pm \Ga,\pm \Ga)$ arises. 

We should stress that there is nothing special about the above construction;
any configuration in $\Spin(2l)$ which is such that it satisifies the 
right relations and breaks the $\so(2l)$ symmetry completely will work as a basis for the construction of a component of the moduli space.

For later purposes it is useful to reduce the above $\Pin(2)$ expressions to $\mathrm{O}(2)$ matrices 
\be \label{SOfrac2}
\column{-i \si_y}{\si_z}{\si_z}, \;
\column{-i\si_y}{\si_z}{-\si_z}, \;
\column{\si_z}{\si_x}{\si_x}, \;
\column{\si_z}{\si_x}{-\si_x}, \;
\column{\si_z}{i\si_y}{i\si_y}, \;
\column{\si_z}{i\si_y}{-i\si_y} .
\ee
Using these building blocks instead will lead to holonomies that belong to $\SO$ groups rather than $\Spin$ groups.

\subsubsection{The  $m = (\id, \Ga, \Ga)$ and $m = (\id, -\Ga, -\Ga)$ components for $G = \Spin (4 k)$}
For the ${\cal M}_k$ and ${\cal M}_k^\prime$ components, there is no prefactor involving the building blocks in  (\ref{Spinfrac2}). The holonomies belong to a maximal torus of $\Spin(2 k)$ and can be written in terms of gamma matrices. However, it will be more convenient to use an $\left[\SU (2) \times \SU (2) \right]^k / \sim$ subgroup of $\Spin (4 k)$, where the equivalence relation identifies the elements $(-\id, -\id)$ of the $k$ factors, and write the holonomies as  
\beq \label{storus}
\column{U_1}{U_2}{U_3} = \column{s_1}{s_2}{s_3} \equiv  \left[ \column{A}{B}{\pm B} , \column{t_1}{t_2}{t_3} \right] \left[ \column{A}{B}{B} , \column{t_1}{t_2}{t_3} \right]^{k-1} .
\eeq
Here the fixed $\SU (2)$ elements $A$ and $B$ obey $A B = - B A$ (e.g.~$A=i\si_z$ and $B=i\si_x$) and each $t_i$ belong to a maximal torus of $\SU (2)$. The signs label the two components. Note that the $\pm$ signs can be moved to any of the other $k-1$ factors, and thus the above expression is symmetric under permutation of the $k$ factors.

Enhanced $\spl (2 n_1) \oplus \spl (2 n_2) \oplus \so (2 n_3) \oplus \ldots \oplus \so (2 n_8)$ symmetry occurs when
\bea \label{torus2}
&&\column{t_1}{t_2}{t_3}^k  =   \\ 
&& \!\!\!\!\! \col{\id}{\id}{\id}^{n_1} \col{\id}{\id}{\!\!-\id\!\!}^{n_2} 
 \col{\id}{i \si_3}{\i \si_3}^{n_3} \col{\id}{i \si_3}{\!\!-i\si_3\!\!}^{n_4} \col{i \si_3}{\id}{\id}^{n_5} \col{i \si_3}{\id}{-\id}^{n_6} \col{i \si_3}{i \si_3}{i\si_3}^{n_7} \col{i \si_3}{i \si_3}{\!\!-i\si_3\!\!}^{n_8} \non \!\!.
\eea
These eight possibilities can be visualised as the corners of a cube.
Why the unbroken gauge symmetry is precisely as above can be understood as follows: 
The $(8 k^2 - 2 k)$-dimensional adjoint representation of $\Spin (4 k)$ decomposes under $[\SU(2)\times \SU(2)]\times \cdots \times [\SU(2)\times \SU(2)]$ as
\beq
k \left[ (\ldots, 3, 1, \ldots) \oplus (\ldots, 1, 3, \ldots) \right] \oplus \frac{1}{2} k (k - 1) (\ldots, 2, 2, \ldots, 2, 2, \ldots) .
\eeq
The $(3,1)$ generators are always broken. The $(1,3)$ generators are unbroken for $n_1$ and $n_2$, and broken to to a single generator for the other $n_i$. The spectrum of $A\otimes A$ and $B\otimes B$ 
in the $(2, 2)$ representation of $\SU (2) \times \SU (2)$ is easily determined to be:
\beq
\left(
\begin{array}{c}
A \otimes A \cr
B \otimes B
\end{array}
\right) \in \left\{
\begin{array}{cccc}
1 & 1 & - 1 & -1 \cr
1 & -1 & 1 & -1
\end{array}
\right \} .
\eeq
from which one can deduce  the number of unbroken generators coming from the $(\ldots,2,2,\ldots,2,2,\ldots)$ pieces.

From the expressions for the holonomies given earlier, we see that the center element $\Ga = [(-1, 1)]^k$ acts trivially in the 1 direction. It acts non-trivially and equally in the 2- and 3-directions, unless the two planes are equal (i.e.~$n_{2l-1}=n_{2l}$). The center element $-\id = ((-\id, -\id), [(\id, \id)]^{k - 1})$ acts non-trivially in the 1-direction if and only if $n_5 = n_6 = n_7=n_8= 0$ (similar statements hold for the other directions). In this way, one finds the contributions
\bea
f_k ((\id, \Ga, \Ga), (\id, \id, \id)) & = & \frac{1}{2} Q^2 (q^2) P_{\mathrm{even}}^6 (q^2) + \frac{1}{2} Q (q^4) P_{\mathrm{even}}^3 (q^4)  \cr
f_k ((\id, \Ga, \Ga), (\id, -\id, -\id)) & = & \frac{1}{2} Q^2 (q^2) P_{\mathrm{even}}^6 (q^2) - \frac{1}{2} Q (q^4) P_{\mathrm{even}}^3 (q^4) \cr
f_k ((\id, \Ga, \Ga), (\id, \Ga, \Ga)) & = & \frac{1}{2} Q^2 (q^2) P_{\mathrm{even}}^6 (q^2) + \frac{1}{2} Q (q^4) P_{\mathrm{even}}^3 (q^4) \cr
f_k ((\id, \Ga, \Ga), (\id, -\Ga, -\Ga)) & = & \frac{1}{2} Q^2 (q^2) P_{\mathrm{even}}^6 (q^2) - \frac{1}{2} Q (q^4) P_{\mathrm{even}}^3 (q^4) \cr
f_k ((\id, \Ga, \Ga), (\Ga, \id, \id)) & = & \frac{1}{2} Q^2 (q^2) P_{\mathrm{even}}^2 (q^2) + \frac{1}{2} Q (q^4) P_{\mathrm{even}} (q^4) \cr
f_k ((\id, \Ga, \Ga), (\Ga, -\id, -\id)) & = & \frac{1}{2} Q^2 (q^2) P_{\mathrm{even}}^2 (q^2) - \frac{1}{2} Q (q^4) P_{\mathrm{even}} (q^4) \cr
f_k ((\id, \Ga, \Ga), (\Ga, \Ga, \Ga)) & = & \frac{1}{2} Q^2 (q^2) P_{\mathrm{even}}^2 (q^2) + \frac{1}{2} Q (q^4) P_{\mathrm{even}} (q^4) \cr
f_k ((\id, \Ga, \Ga), (\Ga, -\Ga, -\Ga)) & = & \frac{1}{2} Q^2 (q^2) P_{\mathrm{even}}^2 (q^2) - \frac{1}{2} Q (q^4) P_{\mathrm{even}} (q^4) . 
\eea

For the ${\cal M}_{k - 3}$ and ${\cal M}_{k - 3}^\prime$ components, the holonomies are
\beq \label{s4m3}
\column{U_1}{U_2}{U_3} = \column{ \half(1+\ga_1\ga_2)(1+\ga_3\ga_4)\ga_6 \ga_8    \ga_{10}  \ga_{12} }{ 
{\ts \frac{1}{4}} \ga_2\ga_4(\ga_5 - \ga_6)(\ga_7 - \ga_8) (1-\ga_9\ga_{10})(1-\ga_{11}\ga_{12}) }{
{\ts \frac{1}{4}}\ga_2 \ga_3 (\ga_5 - \ga_6)(\ga_7 + \ga_8)(1-\ga_9\ga_{10})(1+\ga_{11}\ga_{12})} \column{s_1}{s_2}{s_3}  ,
\eeq
where $(s_1,s_2,s_3)$ is of the same form as in (\ref{storus}). To find the enhanced symmetry we again look at the spectrum.  The $(8 k^2 - 2 k)$-dimensional adjoint representation of $\Spin (4 k)$ decomposes under $\Spin(12)\times[\SU(2)\times \SU(2)]\times \cdots \times [\SU(2)\times \SU(2)]\Spin(4k-12)$ as
\bea
&& (66,\ldots) \oplus (k - 3) \left[ (1, \ldots, 3, 1, \ldots) \oplus (1, \ldots, 1, 3, \ldots) \right] \cr
&& \oplus (k - 3) (12, \ldots, 2, 2, \ldots) \oplus \frac{1}{2} (k - 3) (k - 4) (\ldots, 2, 2, \ldots, 2, 2, \ldots) .
\eea
In addition to the above results, one also needs to determine spectrum 
of the prefactor in (\ref{s4m3}) tensored with the triplet $(A,B,B)$ in the
$(12,2)$ representation of $\Spin (12) \times \SU (2)$. Since we need the vector representation, it is convenient to reduce the $\Spin(12)$ expression to $\SO(12)$ using (\ref{SOfrac2}). The result of the calculation is that enhanced $\spl (2 n_1) \oplus \spl (2 n_2) \oplus \so (2 n_3 + 1) \oplus \ldots \oplus \so (2 n_8 + 1)$ symmetry occurs when the $t_i$'s take the values in (\ref{torus2}).

The analysis for these components is analogous to the one for the 
preceeding cases, with the difference that all center elements ($-\id$, $\Ga$ and $-\Ga$) act trivially in the 1-direction.  The resulting contributions are:
\bea
f_{k-3} ((\id, \Ga, \Ga), (\id, \id, \id)) & = & \frac{1}{2} Q^2 (q^2) P_{\mathrm{odd}}^6 (q^2) + \frac{1}{2} Q (q^4) P_{\mathrm{odd}}^3 (q^4)  \cr
f_{k-3} ((\id, \Ga, \Ga), (\id, -\id, -\id)) & = & \frac{1}{2} Q^2 (q^2) P_{\mathrm{odd}}^6 (q^2) - \frac{1}{2} Q (q^4) P_{\mathrm{odd}}^3 (q^4) \cr
f_{k-3} ((\id, \Ga, \Ga), (\id, \Ga, \Ga)) & = & \frac{1}{2} Q^2 (q^2) P_{\mathrm{odd}}^6 (q^2) + \frac{1}{2} Q (q^4) P_{\mathrm{odd}}^3 (q^4) \cr
f_{k-3} ((\id, \Ga, \Ga), (\id, -\Ga, -\Ga)) & = & \frac{1}{2} Q^2 (q^2) P_{\mathrm{odd}}^6 (q^2) - \frac{1}{2} Q (q^4) P_{\mathrm{odd}}^3 (q^4) \,.
\eea
The total partition functions for these values of $m$ are given by $f (m, e) = f_k (m, e) + f_{k - 3} (m, e)$.

\subsubsection{The $m = (\id, \Ga, \Ga)$ components for $G = \Spin (4 k + 2)$}
For the ${\cal M}_{k - 1}^{(c)}$, $c = 0, 1, 2, 3$ components, the holonomies are 
\beq
\column{U_1}{U_2}{U_3} = \column{\half(1+\ga_1\ga_2)\ga_4\ga_6}{\half\ga_2(\ga_3-\ga_4)(1-\ga_5\ga_6)}{\half\ga_2(\ga_3-\ga_4)(1-\ga_5\ga_6)} \column{s_1}{s_2}{s_3} \!,
\eeq
and
\beq
\column{U_1}{U_2}{U_3} = \column{\half(1+\ga_1\ga_2)\ga_4\ga_6}{\half\ga_2(\ga_3-\ga_4)(1-\ga_5\ga_6)}{\half\ga_1(\ga_3+\ga_4)(1+\ga_5\ga_6)}  \column{s_1}{s_2}{s_3} \!,
\eeq
where again $(s_1,s_2,s_3)$ is of the same form (with a suitable dimension) as in (\ref{storus}). 

The enhanced symmetry is determined as above by using that the $8 k^2 + 6 k + 1$ dimensional adjoint representation of $\Spin (4 k + 2)$ decomposes as
\bea
&& (15, \ldots) \oplus (k - 1) \left[(1, \ldots, 3, 1, \ldots) \oplus (1, \ldots, 1, 3, \ldots) \oplus (6, \ldots, 2, 2, \ldots) \right] \cr
&& \qquad \quad \;\,\, \oplus \, \frac{1}{2} (k - 1) (k -2) \left[(\ldots, 2, 2, \ldots, 2, 2, \ldots) \right] .
\eea
It can be shown that enhanced
\beq
\spl (2 n_1) \oplus \spl(2 n_2) \oplus \so (2 n_3 + 1) \oplus \so (2 n_4 ) \oplus \so (2 n_5 + 1) \oplus \so (2 n_6) \oplus \so (2 n_7+1) \oplus \so (2 n_8) 
\eeq
and
\beq
\spl (2 n_1) \oplus \spl(2 n_2) \oplus \so (2 n_3 ) \oplus \so (2 n_4 +1) \oplus \so (2 n_5 ) \oplus \so (2 n_6+1) \oplus \so (2 n_7) \oplus \so (2 n_8+1) 
\eeq
symmetry respectively occurs when the $t_i$'s take the values in (\ref{torus2}).

On the first holonomy, the centre acts trivially (up to gauge conjugation) and on the second and third it acts equally  and freely (again up to gauge conjugation). From this it follows that 
\bea
f ((\id, \Ga, \Ga), (\id, \id, \id)) & = & Q^2 (q^2) P_{\mathrm{even}}^3 (q^2) P_{\mathrm{odd}}^3 (q^2) \non \\
f ((\id, \Ga, \Ga), (\id, -\id, -\id)) & = & Q^2 (q^2) P_{\mathrm{even}}^3 (q^2) P_{\mathrm{odd}}^3 (q^2)  \\ 
f ((\id, \Ga, \Ga), (\id, \Ga, \Ga)) & = & Q^2 (q^2) P_{\mathrm{even}}^3 (q^2) P_{\mathrm{odd}}^3 (q^2) \non
 \\ 
f ((\id, \Ga, \Ga), (\id, -\Ga, -\Ga)) & = & Q^2 (q^2) P_{\mathrm{even}}^3 (q^2) P_{\mathrm{odd}}^3 (q^2) . \non
\eea

\subsubsection{The $m = (-\id, \Ga, \Ga)$ components for $G = \Spin (4 k)$}
On the ${\cal M}_{k - 1}$ and ${\cal M}_{k - 1}^\prime$ components, the holonomies are 
\beq \label{s4}
\column{U_1}{U_2}{U_3} = \column{ \half(1+\ga_1\ga_2)(1+\ga_3\ga_4)}{\ga_2\ga_4}{\ga_2\ga_3} \column{s_1}{s_2}{s_3}\!,
\eeq
where $(s_1,s_2,s_3)$ is of the form (\ref{storus}). 
Note that there are two other ways to select two columns from (\ref{Spinfrac2}) 
which give values of $m$ in same orbit as the above choice.

By analysing the spectrum as above, one finds that enhanced gauge symmetry 
\beq
\spl (2 n_1) \oplus \spl(2 n_2) \oplus \so (2 n_3 + 1) \oplus \so (2 n_4 +1 ) \oplus \so (2 n_5 ) \oplus \so (2 n_6) \oplus \so (2 n_7) \oplus \so (2 n_8) 
\eeq
occurs when the $t_i$'s take the values in (\ref{torus2}).

On the ${\cal M}_{k - 2}$ and ${\cal M}_{k - 2}^\prime$ components, the holonomies are 
\beq \label{s8}
\column{U_1}{U_2}{U_3} = \column{\ga_2\ga_4\ga_6\ga_8}{{\ts \frac{1}{4}}(\ga_1-\ga_2)(\ga_3-\ga_4)(1-\ga_5\ga_6)(1-\ga_7\ga_8)}{ {\ts \frac{1}{4}}(\ga_1-\ga_2)(\ga_3+\ga_4)(1-\ga_5\ga_6)(1+\ga_7\ga_8)} \column{s_1}{s_2}{s_3}\!,
\eeq
where $(s_1,s_2,s_3)$ is of the form (\ref{storus}). 
Again there are two other ways to select four columns 
from (\ref{Spinfrac2}) which give values of $m$ in the same orbit 
as the above choice. Note that (\ref{s8}) is the `complement' of (\ref{s4}), cf.~\cite{Keurentjes:2000}.

One finds that enhanced gauge symmetry 
\beq
\spl (2 n_1) \oplus \spl(2 n_2) \oplus \so (2 n_3 ) \oplus \so (2 n_4 ) \oplus \so (2 n_5 +1) \oplus \so (2 n_6+1) \oplus \so (2 n_7+1) \oplus \so (2 n_8+1) 
\eeq
occurs when the $t_i$'s take the values in (\ref{torus2}).

Eight of the $64$ large gauge transformations may be compensated by conjugation with $g=U_1^kU_2^lU_3^m$ for $k,l,m=0,1$, so only $64 / 8 = 8$ different values of $e$ appear. The corresponding partition functions are given  below
\bea
f ((-\id, \Ga, \Ga), (\id, \id, \id)) &\!\!\! = & \!\!\!\frac{1}{2} Q^2 (q^2)[ P_{\mathrm{odd}}^2  (q^2) P_{\mathrm{even}}^4(q^2) +  P_{\mathrm{odd}}^4  (q^2) P_{\mathrm{even}}^2(q^2)] \non \\
&& \!\!\!+\, \frac{1}{2} Q (q^4) [P_{\mathrm{odd}}(q^4) P_{\mathrm{even}}^2(q^4) +  P_{\mathrm{odd}}^2(q^4) P_{\mathrm{even}}(q^4)] \non \\
f ((-\id, \Ga, \Ga), (-\id, \Ga, \Ga)) &\!\!\! = & \!\!\!\frac{1}{2} Q^2 (q^2)[ P_{\mathrm{odd}}^2  (q^2) P_{\mathrm{even}}^4(q^2) + P_{\mathrm{odd}}^4  (q^2) P_{\mathrm{even}}^2(q^2)] \non \\
&& \!\!\!+\, \frac{1}{2} Q (q^4) [P_{\mathrm{odd}}(q^4) P_{\mathrm{even}}^2(q^4) +  P_{\mathrm{odd}}^2(q^4) P_{\mathrm{even}}(q^4)] \non \\
f ((-\id, \Ga, \Ga), (\id, -\id, -\id)) & \!\!\!= & \!\!\! \frac{1}{2} Q^2 (q^2) [P_{\mathrm{odd}}^2  (q^2) P_{\mathrm{even}}^4(q^2) +  P_{\mathrm{odd}}^4  (q^2) P_{\mathrm{even}}^2(q^2)] \non \\
&& \!\!\!-\, \frac{1}{2} Q (q^4) [P_{\mathrm{odd}}(q^4) P_{\mathrm{even}}^2(q^4) - P_{\mathrm{odd}}^2(q^4) P_{\mathrm{even}}(q^4)] \non \\
f ((-\id, \Ga, \Ga), (-\id, -\Ga, -\Ga)) 
& \!\!\!= & \!\!\! \frac{1}{2} Q^2 (q^2) [P_{\mathrm{odd}}^2  (q^2) P_{\mathrm{even}}^4(q^2) +  P_{\mathrm{odd}}^4  (q^2) P_{\mathrm{even}}^2(q^2)] \non \\
&& \!\!\!-\, \frac{1}{2} Q (q^4)[ P_{\mathrm{odd}}(q^4) P_{\mathrm{even}}^2(q^4) - P_{\mathrm{odd}}^2(q^4) P_{\mathrm{even}}(q^4)] \non \\
f ((-\id, \Ga, \Ga), (\Ga,\id, \id)) 
&\!\!\! = &\!\!\! \frac{1}{2} Q^2 (q^2) P_{\mathrm{odd}}^2  (q^2) + \frac{1}{2} Q(q^4) P_{\mathrm{odd}}(q^4)  \non \\
f ((-\id, \Ga, \Ga), (-\Ga, \Ga, \Ga)) & \!\!\!= & \!\!\! \frac{1}{2} Q^2 (q^2) P_{\mathrm{odd}}^2  (q^2) + \frac{1}{2} Q(q^4) P_{\mathrm{odd}}(q^4)   \\
f ((-\id, \Ga, \Ga), (\Ga, -\id, -\id)) & \!\!\!= & \!\!\! \frac{1}{2} Q^2 (q^2) P_{\mathrm{odd}}^2  (q^2) - \frac{1}{2} Q(q^4) P_{\mathrm{odd}}(q^4)  \non \\
f ((-\id, \Ga, \Ga), (-\Ga, -\Ga, -\Ga)) & \!\!\!= &\!\!\! \frac{1}{2} Q^2 (q^2) P_{\mathrm{odd}}^2  (q^2) - \frac{1}{2} Q(q^4) P_{\mathrm{odd}}(q^4) . \non
\eea

\subsection{Orientifold interpretation}
As discussed in \cite{Witten:1998,Keurentjes:2000,Henningson:2007} the above moduli spaces can be described in terms of orientifolds. This language is convenient since it immediately gives the unbroken gauge symmetry. When the components of $m$ are $\id$ or $-\id$, the relevant orientifold contains eight $O^-$ orientifold planes. As usual, $n$ D-branes located at one of these $O^-$ planes leads to $\so(n)$ gauge enhancement. A single D-brane stuck at one of the eight $O^{-}$ orientifold planes corresponds to the eight building blocks in (\ref{Spinfrac}). Thus the prefactors contained in $\Spin(l)$ correspond to configurations of stuck (`fractional') branes. In addition, each of the the parameters in (\ref{torus}) correspond to the location of a brane-mirror pair. 
When the components of $m$ contains at least one of $\Ga$ or $-\Ga$, the relevant orientifold contains two $O^+$ and six $O^-$ planes \cite{Keurentjes:2000}, where $n$ D-branes located at one of the $O^+$ planes leads to $\spl(n)$ gauge enhancement. Only the $O^-$ planes can support an odd number of branes, and a single stuck D-brane at one of six $O^{-}$ planes corresponds to the six building blocks in (\ref{Spinfrac2}). The $\Spin(2l)$ prefactors (zero-rank triples) are again constructed from the stuck branes. The part in (\ref{storus}) describes brane-mirror pairs. Note that the total number of branes in the 2$O^+$ 6 $O^-$ orientifold is only half the number in the 8 $O^-$ orientifolds.

\subsection{$S$-duality}
$S$-duality is the statement that the number of bound states with quantum numbers 
$(m,e)$ should agree with the number of states with $(e,m^{-1})$ (we write $m^{-1}$ rather than $-m$ since we are using multiplicative notation). Looking at the tables in the previous section we can check if the spectrum of bound states is $S$-dual (we occasionally also need to take the $\SL(3,\ZZ)$ symmetry into account). We will need some identities for the generating functions:
\bea
Q(q^2)P(q^2) = Q(q) \label{pqid} \\
P(-q)P(q) = P(-q^2) \label{ppid} \\
Q(q) P(-q^2) = 1  ,  \label{euler}
\eea
where the last line is Euler's famous identity which is easy to prove. The theta functions with zero argument (theta constants) can be written in terms of infinite products as 
\bea \label{thetaexp}
\theta_2(q) &=& 2q^{1/4} \prod_{k=1}^{\infty} (1-q^{2k})(1+q^{2k})^2 \,,\non \\
 \theta_3(q) &=&  \prod_{k=1}^{\infty} (1-q^{2k})(1+q^{2k-1})^2 \,, \\
\theta_4(q) &=& \prod_{k=1}^{\infty} (1-q^{2k})(1-q^{2k-1})^2 \,. \non
\eea
The theta constants satisfy the following identities (these are not all independent and are essentially all identities of this type):
\bea
\theta_2(q)^4 &= &\theta_3(q)^4 - \theta_4(q)^4 \,, \label{eid1} \\ 
2\,\theta_2(q^2)^2 &=& \theta_3(q)^2 - \theta_4(q)^2\,, \\
2\,\theta_2(q^4) &=& \theta_3(q) - \theta_4(q)\,, \label{eid3} \\
2\,\theta_3(q^2)^2& =&  \theta_3(q)^2 + \theta_4(q)^2\,, \label{id1} \\
2\, \theta_3 (q^4)& =& \theta_3(q) + \theta_4(q)\,, \label{id2} \\
\theta_4(q^2)^2 &=& \theta_3(q)\theta_4(q) \, . \label{id3} 
\eea

We will only check $S$-duality for the non-trivial cases, and take into account the symmetry between $\Ga$ and $-\Ga$ that corresponds to interchanging the two spinor representations. For the number of states with $(m,e)=((\id,\id, \id),(\id,-\id,-\id))$ to agree with the dual number we require 
\be
{\ts \frac{1}{8}} P_{\mathrm{odd}}^8 (q) + {\ts \frac{1}{8}} P_{\mathrm{even}}^8 (q) - {\ts \frac{1}{8}} P_{\mathrm{odd}}^4 (q^2) - {\ts \frac{1}{8}} P_{\mathrm{even}}^4 (q^2) = {\ts \frac{1}{4}} P_{\mathrm{even}}^4 (q) P_{\mathrm{odd}}^4 (q) + {\ts \frac{3}{4}} P_{\mathrm{even}}^2 (q^2) P_{\mathrm{odd}}^2 (q^2) , 
\ee
which can be rewritten as
\be
P(q)^6 P(-q)^2 + P(-q)^6 P(q)^2  + 2 P(q)^4 P(-q)^4 
 = 2[ P(q^2)^4 + P(-q^2)^4] ,
\ee 
which in turn is equivalent to (\ref{id1}), using (\ref{ppid}).

For the number of states with $(m,e)=((\id,\Ga,\Ga),(\id,\id,\id))$ to agree with the dual number we require 
\bea
&&\frac{1}{2} Q(q^2)^2 [P_{\mathrm{odd}}(q^2)^6 + P_{\mathrm{even}}(q^2)^6+ 2 P_{\mathrm{even}}^3 (q^2) P_{\mathrm{odd}}^3 (q^2) ] \\&&+ \frac{1}{2} Q(q^4) [P_{\mathrm{odd}}(q^4)^3 + P_{\mathrm{even}}(q^4)^3 ] \non =
\frac{1}{4} P_{\mathrm{even}}^4 (q)+\frac{3}{4} P_{\mathrm{even}}^2 (q^2) .
\eea
A similar analysis for $(m,e)=((1,\Ga,\Ga),(1,-1,-1))$ leads to
\bea
&&\frac{1}{2} Q(q^2)^2 [P_{\mathrm{odd}}(q^2)^6 + P_{\mathrm{even}}(q^2)^6+ 2 P_{\mathrm{even}}^3 (q^2) P_{\mathrm{odd}}^3 (q^2) ] \\ &&- \frac{1}{2} Q(q^4) [P_{\mathrm{odd}}(q^4)^3 + P_{\mathrm{even}}(q^4)^3 ] \non =
\frac{1}{4} P_{\mathrm{odd}}^4 (q)+\frac{3}{4} P_{\mathrm{odd}}^2 (q^2) .
\eea
The difference of the above two equations can be proven using (\ref{pqid}), (\ref{euler}) together with (\ref{id2}). The sum can be proven using (\ref{pqid})--(\ref{euler}) together with (\ref{id1}) (the identity (\ref{id3}) is also useful).

For $(m,e)=((-\id,\Ga,\Ga),(\id,\id,\id))$ the $S$-duality requirement is
\bea
&&\frac{1}{2} Q^2(q^2) [P^2_{\mathrm{odd}}(q^2) P^4_{\mathrm{even}}(q^2) + P^4_{\mathrm{odd}}(q^2) P^2_{\mathrm{even}}(q^2) + 2 P^3_{\mathrm{odd}}(q^2) P^3_{\mathrm{even}}(q^2)]  \\
&& + \frac{1}{2} Q(q^4) [P_{\mathrm{odd}}(q^4) P^2_{\mathrm{even}}(q^4) + P^2_{\mathrm{odd}}(q^4) P_{\mathrm{even}}(q^4)]
= \frac{1}{4} P_{\mathrm{even}}^4 (q)-\frac{1}{4} P_{\mathrm{even}}^2 (q^2) . \non
\eea
Similarly, for $(m,e)=((-\id,\Ga,\Ga),(\id,-\id,-\id))$ we need
\bea
&&\frac{1}{2} Q^2(q^2) [P^2_{\mathrm{odd}}(q^2) P^4_{\mathrm{even}}(q^2) + P^4_{\mathrm{odd}}(q^2) P^2_{\mathrm{even}}(q^2)+ 2 P^3_{\mathrm{odd}}(q^2) P^3_{\mathrm{even}}(q^2) ]  \\
&& - \frac{1}{2} Q(q^4) [P_{\mathrm{odd}}(q^4) P^2_{\mathrm{even}}(q^4) + P^2_{\mathrm{odd}}(q^4) P_{\mathrm{even}}(q^4)] =
\frac{1}{4} P_{\mathrm{odd}}^4 (q) - \frac{1}{4} P_{\mathrm{odd}}^2 (q^2) . \non
\eea
The difference of the above two equations can again be proven using (\ref{pqid})--(\ref{euler}) together with (\ref{id2}). The sum can be proven using (\ref{pqid})--(\ref{euler}) together with (\ref{id1}) (the identity (\ref{id3}) is also useful).

Finally, $S$-duality for $(e,m) = ( (\id,\Ga,\Ga), (-\id,\Ga,-\Ga) )$ only needs to be checked for $\Spin(4k{+}2)$ and requires 
\be
 \half Q^2(q^2) P_{\mathrm{even}}(q^2)^2 - \half Q(q^4) P_{\mathrm{even}}(q^4) = \half Q^2(q^2) P_{\mathrm{odd}}(q^2)^2 + \half Q(q^4) P_{\mathrm{odd}}(q^4) ,
\ee
 which is equivalent to 
\be
Q^2(q^2) P(q^2)P(-q^2) = Q(q^4)P(q^4) ,
\ee
which in turn  follows from (\ref{pqid})--(\ref{euler}). 

Above we only analyzed the $S$ transformation of the $S$-duality group (\ref{S}); the $T$ transformation of the $S$-duality group acts as $(m,e)\rar(m,em)$ (again we use multiplicative notation). 
As a perusal of the tables above show, $T$ is also a symmetry. 
To conclude, we have seen that $S$ duality is valid provided that the number of bound states of supersymmetric matrix quantum mechanics agree with those listed in the introduction. (It is not difficult to convince oneself that this is the unique solution, cf.~\cite{Henningson:2007}.) Note that the same conclusion was obtained in our previous paper \cite{Henningson:2007} by considering the $G=\Spin(2n{+}1)$ and $G=\Sp(2n)$ $S$-dual theories. However, for those cases it was the identities (\ref{eid1})--(\ref{eid3}) that were relevant whereas in this paper (\ref{id1})--(\ref{id3}) were used.

\setcounter{equation}{0}
\section{The exceptional groups}
The remaining cases, i.e. $G \simeq G_2, F_4, E_6, E_7, E_8$, may be analyzed as follows: Each component of the moduli space of flat connections may be described by finding a suitable subgroup $K {\times} H \subset G$, where $K$ admits an almost commuting triple $(k_1, k_2, k_3)$ of the appropriate magnetic 't Hooft flux $m$, and $H$ is simple. The holonomies are then given by
\beq
\left(
\begin{array}{l}
U_1 \cr
U_2 \cr
U_3
\end{array}
\right) =
\left(
\begin{array}{l}
k_1, t_1 \cr
k_2, t_2 \cr
k_3, t_3
\end{array}
\right) ,
\eeq
 where the $t_i$, $i = 1, 2, 3$ belong to a maximal torus $T$ of $H$. The first step is to classify all possible semi-simple subgroups of $H$. Such a subgroup $S$ is unbroken precisely when the $t_i$ are elements of the center of $S$. We are only interested in equivalence classes of such choices modulo conjugation. Conjugation by elements of $H$ corresponds to the Weyl group of $H$; to take conjugation by arbitrary elements of $G$ into account, we must also divide by those automorphisms  of $H$ that leave the fundamental representation of $G$ invariant. One should then determine the unbroken subalgebra $s$ of the Lie algebra of $G$ in these cases. In most cases, $s$ is the Lie algebra of $S$, but in some cases it is larger, because generators of the Lie algebra of $G$ that do not belong to the Lie algebra of $K {\times} H$ may be unbroken. As before, such a configuration contributes $\dim V_s$ states. Finally, one must investigate the transformation properties of these states under large gauge transformations, which act by multiplication of the holonomies $U_i$ by elements of the center $C$ of $G$, to determine their values of the electric 't Hooft flux $e$.

\vbox{

\subsection{$G \simeq G_2$}
This group has $g^\vee = 4$ and a trivial center $C \simeq 1$. 

The moduli space contains a  $2$-dimensional component, for which $K$ is trivial and $H \simeq G_2$. The possible semi-simple subgroups of $H$ are $G_2$, $\SU (3)$, and $\SU (2) \otimes \SU (2) / \sim$, where $\sim$ denotes the equivalence relation $(-\id_2, -\id_2) \sim (\id_2, \id_2)$. $S \simeq G_2$ is unbroken when the $t_i$ belong to the trivial center of $G_2$, and this single configuration contributes $\dim V_{G_2} = 2$ states (assuming that $\Delta_{G_2} = 0$). $S \simeq SU (3)$ is unbroken when the $t_i$ belong to the $\mathbb Z_3$ center of $SU (3)$. Of these $27$ choices, the one in which all the $t_i$ equal the unit element actually has $G_2$ symmetry and should not be taken into account. The remaining $26$ are pairwise equivalent under the complex conjugation automorphism of $\SU (3)$, so there are $13$ inequivalent configurations, each of which contributes $\dim V_{\su (3)} = 1$ each. Similarly, $S \simeq SU (2) \times SU (2) / \sim$ is unbroken for $8$ different choices of the $t_i$, one of which actually gives unbroken $G_2$ and should be removed, whereas the remaining $7$ gives $\dim V_{\su (2)} \times \dim V_{\su (2)} = 1$ each.

The moduli space also contains a $0$-dimensional component, for which $K \simeq G_2$ and $H$ is trivial. This gives $1$ state. All together, we get 
\beq
\begin{array}{llr}
\underline{K} & \underline{s} & \underline{\mathrm{states}} \cr
1 & G_2 & 2 \times 1 \cr
   & \su (3) & 13 \cr
   & \su (2) \oplus \su (2) & 7 \cr
\cr
G_2 & \emptyset & 1 \cr
& & \underline{\;\;\;\;} \cr
& & 23
\end{array} 
\eeq

\subsection{$G \simeq F_4$}
This case, with $g^\vee = 9$ and $C \simeq 1$, is rather similar to the $G_2$ case, and we will only display the results:
\beq
{\scriptsize
\begin{array}{llr}
\underline{K} & \underline{s} & \underline{\mathrm{states}}  \cr
1 & F_4 & 4 \times 1 \cr
   & \so (9) &  2 \times 7 \cr
   & \su (3) \oplus \su (3) & 13 \cr
   & \spl (6) \oplus \su (2) & 2 \times 7 \cr
   & \so (8) & 2 \times 7 \cr
   & \so (6) \oplus \so (3) &  28 \cr
   & \spl (4) \oplus \su (2) \oplus \su (2) & 21 \cr
   & \su (2) \oplus \su (2) \oplus \su (2) \oplus \su (2) & 7 \cr
\cr
G_2 & \su (2) & 1 \cr
\cr
F_4 & \emptyset & 1 \cr
\cr
F_4 & \emptyset & 1 \cr
& & \underline{\;\;\;\;} \cr
& & 118
\end{array}
}
\eeq

}

\newpage 

\subsection{$G \simeq E_6$}
This group  (with $g^\vee = 12$) has a non-trivial center $C \simeq \mathbb Z_3$, so we must distinguish the cases with different values of $m$, and also determine the values of $e$. 

When $m$ is trivial, the results are
\beq
{\scriptsize
\begin{array}{llrrr}
\underline{K} & \underline{s} & \underline{\mathrm{states}} & \underline{e \; {\rm trivial}} & \underline{e \; {\rm non{-}trivial}} \cr
1 & E_6 & 3 \times 27 & 3 & 3 \cr
   & \su (6) \oplus \su (2) & 189 & 7 & 7 \cr
   & \su (3) \oplus \su (3) \oplus \su (3) & 234 & 26 & 8 \cr
\cr
G_2 & \su (3) & 27 & 1 & 1 \cr
\cr
E_6 & \emptyset & 1 & 1 & 0 \cr
\cr
E_6 & \emptyset & 1 & 1 & 0 \cr
& & & \underline{\;\;\;\;} & \underline{\;\;\;\;} \cr
& & & 39 & 19
\end{array}
}
\eeq
In most of these cases, $C$ acts non-trivially on all three holonomies, so that there is $1/27$ of the total number of states for each of the single trivial and the $26$ non-trivial values of $e$. The exception is the $\su (3) \oplus \su (3) \oplus \su (3)$ states, where $C$ acts trivially on one of the holonomies, so that $1/9$ of the total number of states has $e$ trivial, and the remaining states are equally divided between the non-trivial values of $e$.
 
When $m$ is non-trivial, there are three copies of each component corresponding to the possible values of $e$. $C$ acts freely on these components in the direction of $m$ and trivially in the other two directions, so there will be an equal number of states for each value of $e$ parallel to $m$, i.e. $e$ is trivial, equal to $m$ or the inverse of $m$. One set of components are obtained by taking $K  \times H \simeq \SU (3) \times G_2 \subset E_6$ with the branching of the adjoint representation 
\beq
78 = (8, 1) \oplus (1, 14) \oplus (8, 7) .
\eeq
Just as in the $G \simeq G_2$ case considered above, the possible unbroken Lie algebras are $s \simeq G_2, \su (3), \su (2) \oplus \su (2)$. In the $s \simeq G_2$ and $s \simeq \su (2) \oplus \su (2)$ cases, the unbroken generators are given by a subset of the $(1, 14)$ representation, and as before get $2 \times 1$ and $7$ states respectively. The $s \simeq \su (3)$ case is different though: For  $4$ out of the $13$ configurations in which the holonomy in the direction parallel to $m$ is trivial, $6$ of the generators of the $(8, 7)$ representation are unbroken in addition to the $8$ $\su (3)$ generators in $(1, 14)$. Together these generate an unbroken $G_2$ algebra (which is related by conjugation in $E_6$ to the 'standard' algebra $H \simeq G_2$), so these configurations  should not be counted. Adding also the contributions from the set of components with $K \simeq E_6$ and $H$ trivial, we get
\beq
{\scriptsize
\begin{array}{llr}
\underline{K} & \underline{s} & \underline{\mathrm{states}} \cr
\SU (3) & G_2 & 2 \times 1 \cr
& \su (3) & 9 \cr
& \su (2) \oplus \su (2) & 7\cr
\cr
E_6 & \emptyset & 1 \cr
& & \underline{\;\;\;\;} \cr
& & 19
\end{array}
}
\eeq
The appearance of $19$ states both for $m$ trivial, $e$ non-trivial and for $m$ non-trivial, $e$ parallel to $m$ is a manifestation of $S$-duality.

\subsection{$G \simeq E_7$}
This case has $g^\vee = 18$, $C \simeq \mathbb Z_2$, and is rather similar to the previous one. For 
$m$ trivial, we get
\beq
{\scriptsize
\begin{array}{llrrr}
\underline{K} & \underline{s} & \underline{\mathrm{states}} & \underline{e \; {\rm trivial}} & \underline{e \; {\rm non{-}trivial}} \cr
1 & E_7 & 6 \times 8 & 6 & 6 \cr
   & \su (8) & 28 & 7 & 3 \cr
   & \su (6) \oplus \su (3) & 104 & 13 & 13 \cr
   & \so (12) \oplus \su (2) & 3 \times 56 & 21 & 21 \cr
   & \so (8) \oplus \su (2) \oplus \su (2) \oplus \su (2) & 2 \times 56 & 14 & 14 \cr
   & \so (6) \oplus \so (6) \oplus \su (2) & 112 & 14 & 14 \cr
   & \su (2)^7 & 8 & 1 & 1 \cr
\cr
G_2 & \spl (6) & 2 \times 8 & 2 & 2 \cr
\cr
E_6 & \su (2) & 8 & 1 & 1 \cr
\cr
E_6 & \su (2) & 8 & 1 & 1 \cr
\cr
E_7 & \emptyset & 1 & 1 & 0 \cr
\cr
E_7 & \emptyset & 1 & 1 & 0 \cr
& & & \underline{\;\;\;\;} & \underline{\;\;\;\;} \cr
& & & 82 & 76
\end{array}
}
\eeq

For $m$ non-trivial, the components are come in pairs. $C$ acts freely on these components in the direction of $m$ and trivially in the two remaining directions, so there will be an equal number of states for $e$ trivial and $e$ equal to $m$. One set of components is constructed using a $K \times H \simeq \SU (2) \times F_4 \subset E_7$ subgroup under with the branching rule
\beq
133 = (3, 1) \oplus (1, 52) \oplus (3, 26) .
\eeq
The list of possible subgroups of $F_4$ is of course the same as the one presented for the $G \simeq F_4$ case. But their generators, which lie in the $(1, 52)$ representation, may be complemented with generators from the $(3, 26)$ representation and build up larger algebras. A new feature is that some of these algebras are not isomorphic to subalgebras of $F_4$. The spectrum of states with $e$ trivial or with $eÊ= m$ is
\beq
{\scriptsize
\begin{array}{llr}
\underline{K} & \underline{s} & \underline{\mathrm{states}} \cr
A_1 & F_4 & 4 \times 1 \cr
        & \so (9) &  2 \times 4 \cr
        & \su (3) \oplus \su (3) & 13 \cr
        & \spl (6) \oplus \su (2) & 2 \times 4 \cr
        & \spl (8) &  2 \times 3 \cr
        & \so (6) \oplus \so (3) & 16 \cr
        & \so (5) \oplus \so (5) & 6 \cr
        & \so (7) \oplus \su (2) & 12 \cr
\cr
\Spin (12) & \su (2) & 1 \cr
\cr
E_7 & \emptyset & 1 \cr
\cr
E_7 & \emptyset & 1 \cr
& & \underline{\;\;\;\;} \cr
& & 76
\end{array}
}
\eeq
Again, the appearance of $76$ states for $m$ trivial, $e$ non-trivial and for $m$ non-trivial, $e$ parallel to $m$ is a manifestation of $S$-duality.

\subsection{$G \simeq E_8$}
This case, with $g^\vee = 30$ and $C \simeq 1$, is mopre involved but presents no particular new features compared to the previous cases. The spectrum of states is
\beq
{\scriptsize
\begin{array}{llr}
\underline{K} & \underline{s} & \underline{\mathrm{states}} \cr
1 & E_8 & 11 \cr
   & \su (5) \oplus \su (5) & 31 \cr
   & \su (9) & 13 \cr
   & E_7 \oplus \su (2) & 6 \times 7 \cr
   & E_6 \oplus \su (3) & 3 \times 13 \cr
   & \su (6) \oplus \su (3) \oplus \su (2) & 91 \cr
   & \su (3) \oplus \su (3) \oplus \su (3) \oplus \su (3) & 117 \cr
   & \su (8) \oplus \su (2) & 28 \cr
   & \so (16) & 5 \times 7 \cr
   & \so (12) \oplus \su (2) \oplus \su (2) & 3 \times 21 \cr
   & \so (10) \oplus \so (6) & 2 \times 28 \cr
   & \so (8) \oplus \so (8) & 2 \times 2 \times 7 \cr
   & \so (8) \oplus \su (2)^4 & 2 \times 7 \cr
   & \so (6) \oplus \so (6) \oplus \su (2) \oplus \su (2) & 42 \cr
   & \su (2)^8 & 7 \cr
\cr
G_2 & F_4 & 4 \times 1 \cr  
         & \so (7) \oplus \so (3) & 28 \cr
         & \su (3) \oplus \su (3) & 13 \cr
         & \spl (8) & 2 \times 7 \cr
\cr
E_6 & G_2 & 3 \times 1 \cr
        & \su (2) \oplus \su (2) & 7 \cr
        \cr
E_6 & G_2 & 3 \times 1 \cr
        & \su (2) \oplus \su (2) & 7 \cr
\cr
E_7 & \su (2) & 1 \cr
\cr
E_7 & \su (2) & 1 \cr
\cr
E_8 & \emptyset & 1 \cr
\cr
E_8 & \emptyset & 1 \cr
\cr
E_8 & \emptyset & 1 \cr
\cr
E_8 & \emptyset & 1 \cr
\cr
E_8 & \emptyset & 1 \cr
\cr
E_8 & \emptyset & 1 \cr
& & \underline{\;\;\;\;} \cr
& & 704
\end{array}
}
\eeq

\section*{Acknowledgements}

M.H. is a Research Fellow at the Royal Swedish Academy of Sciences.\\
N.W. is supported by a grant from the Swedish Science Council.

\begingroup\raggedright\endgroup

\end{document}